\documentclass{article}
\usepackage{amsmath,amssymb,theorem,mathpazo,color}
\usepackage{graphicx}
\usepackage{url}
\newtheorem{thm}{Theorem}[section]

\newtheorem{prop}[thm]{Proposition}
\newtheorem{defn}[thm]{Definition}

\newtheorem{rem}[thm]{Remark}
\newtheorem{ass}[thm]{Assumption}
\newtheorem{ex}[thm]{Example}
\allowdisplaybreaks
\def\proof {\noindent{\it Proof.}$\quad$$\quad$}
\def\fin   {\hfill{$\Box$}\vspace{5mm}}
\def\l     {\left}
\def\r     {\right}
\def\<     {\langle}
\def\>     {\rangle}

\def\calB  {{\cal B}}
\def\calD  {{\cal D}}
\def\calE  {{\cal E}}
\def\calF  {{\cal F}}
\def\calP  {{\cal P}}

\def\calS  {{\cal S}}

\def\bbD   {{\mathbb D}}
\def\bbE   {{\mathbb E}}
\def\bbF   {{\mathbb F}}

\def\bbN   {{\mathbb N}}
\def\bbP   {{\mathbb P}}
\def\bbQ   {{\mathbb Q}}
\def\bbR   {{\mathbb R}}
\def\ve    {\varepsilon}

\def\vt    {\vartheta}

\def\tP    {\bbP^*}

\def\tN    {\widetilde{N}}
\def\tZ    {\widetilde{Z}}
\def\whK   {\widehat{K}}

\def\tcH   {\widetilde{c}^H}
\def\tvtH  {\widetilde{\vt}^H}
\def\txiH  {\widetilde{\xi}^H}
\def\hvtH  {\widehat{\vt}^H}
\def\hvtHc {\widehat{\vt}^{H-c}}
\def\hvt   {\widehat{\vt}}
\def\hetaH {\widehat{\eta}^H}
\def\hNH   {\widehat{N}^H}
\def\hLH   {\widehat{L}^H}
\def\tQ    {\widetilde{\bbQ}}
\def\hlW   {\widehat{l}^W}
\def\hlN   {\widehat{l}^N}

\begin{document}
\title{A closed-form representation of mean-variance hedging for additive processes via Malliavin calculus}
\author{Takuji Arai\footnote{Department of Economics, Keio University, 2-15-45 Mita, Minato-ku, Tokyo, 108-8345, Japan (arai@econ.keio.ac.jp, tel:+81-3-5427-1411, fax:+81-3-5427-1578)}
        and Yuto Imai\footnote{Department of Mathematics, Waseda University, 3-4-1 Okubo, Shinjyuku-ku, Tokyo 169-8555, Japan (y.imai@aoni.waseda.jp)}}

\maketitle

\begin{abstract}
We focus on mean-variance hedging problem for models whose asset price follows an exponential additive process.
Some representations of mean-variance hedging strategies for jump type models have already been suggested,
but none is suited to develop numerical methods of the values of strategies for any given time up to the maturity.
In this paper, we aim to derive a new explicit closed-form representation, which enables us to develop an efficient numerical method using the fast Fourier transforms.
Note that our representation is described in terms of Malliavin derivatives.
In addition, we illustrate numerical results for exponential L\'evy models.
\end{abstract}

\noindent
{\bf Keywords:} Mean-variance hedging, Additive processes, Malliavin calculus, Fast Fourier transform. \\
{\bf AMS 2010 Subject classification:} 91G20,60H07,91G60.

%
%
\setcounter{equation}{0}
\section{Introduction}
Hedging problem for contingent claims in incomplete markets is a centerpiece of mathematical finance.
Actually, many hedging methods for incomplete markets have been suggested.
Above all, we focus on mean-variance hedging (MVH) problem, which has been studied very well for about three decades.
However, no numerical methods of the values of MVH strategies for any given time up to the maturity for jump type models have been developed,
since any existing representation of MVH strategies is not appropriate for computation.
Thus, we aim to derive a new representation for exponential additive models which makes it possible to develop an efficient numerical method of MVH strategies.
Note that our representation is a closed-form one obtained by means of Malliavin calculus for L\'evy processes.
In addition, we develop a numerical method using the fast Fourier transforms (FFT); and show numerical results for exponential L\'evy models.

We consider throughout an incomplete financial market in which one risky asset and one riskless asset are tradable.
Let $T>0$ be the maturity of our market, and suppose that the interest rate of the riskless asset is $0$ for sake of simplicity.
The risky asset price process, denoted by $S$, is given as a solution to the following stochastic differential equation:
\[
dS_t=S_{t-}\l[\alpha_tdt+\beta_tdW_t+\int_{\bbR_0}\gamma_{t,z}\tN(dt,dz)\r],\hspace{3mm}S_0>0,
\]
where $\bbR_0:=\bbR\setminus\{0\}$, $W$ is a one-dimensional standard Brownian motion, and $\tN$ is the compensated version of a homogeneous Poisson random measure $N$.
Here, $\alpha$ and $\beta$ are deterministic measurable functions on $[0,T]$, and $\gamma$ is also deterministic and jointly measurable on $[0,T]\times\bbR_0$.
We assume $\gamma>-1$, which ensures the positivity of $S$.
Then, $S$ is given as an exponential of an additive process, that is, $\log(S)$ is continuous in probability and has independent increments.
In addition, when $\alpha$ and $\beta$ are given by a real number and a non-negative real number, respectively, and $\gamma_{t,z}=e^z-1$, we call $S$ an exponential L\'evy process.
Let $H$ be a square integrable random variable.
We consider its value as the payoff of a contingent claim at the maturity $T$.
In principle, since our market is incomplete, we cannot find a replicating strategy for $H$, that is, there is no pair $(c,\vt)\in\bbR\times\Theta$ satisfying
\[
H=c+G_T(\vt),
\]
where $\Theta$ is a set of predictable processes, which is considered as the set of all admissible strategies in some sense, and
$G(\vt)$ denotes the gain process induced by $\vt$, that is, $G(\vt):=\int_0^\cdot\vt_udS_u$.
Note that each pair $(c,\vt)\in\bbR\times\Theta$ represents a self-financing strategy.
Instead of finding the replicating strategy, we consider the following minimization problem:
\[
\min_{c\in\bbR, \vt\in\Theta}\bbE\l[\l(H-c-G_T(\vt)\r)^2\r],
\]
and call its solution $(\tcH, \tvtH)\in\bbR\times\Theta$ the MVH strategy for claim $H$ if it exists.
In other words, the MVH strategy is defined as the self-financing strategy minimizing the corresponding $L^2$-hedging error over $\bbR\times\Theta$.
Remark that $\tcH$ gives the initial cost, which is regarded as the corresponding price of $H$;
and $\tvtH_t$ represents the number of shares of the risky asset in the strategy at time $t$.

In addition to MVH strategy, locally risk-minimizing (LRM) strategy has been studied well as alternative hedging method in quadratic way.
Being different from the MVH approach, an LRM strategy is given as a replicating strategy which is not necessarily self-financing.
Thus, we need to take an additional cost process into account.
Roughly speaking, a strategy is said to be LRM if it minimizes in the $L^2$-sense the risk caused by such an additional cost process among all replicating strategies
which are not necessarily self-financing.
For more details, see Schweizer \cite{Sch} and \cite{Sch3}.
As for expressions of LRM strategies, Arai and Suzuki \cite{AS} obtained an explicit form for L\'evy markets using Malliavin calculus for L\'evy processes.
Here, L\'evy market is a similar model framework to ours, but the coefficient functions $\alpha$, $\beta$ and $\gamma$ may have randomness.
In other words, our model is a L\'evy market with deterministic coefficients.

There is much literature on MVH strategies for jump type models.
Among others, Arai \cite{A05}, \v{C}ern\'{y} and Kallsen \cite{CK}, and Jeanblanc et al. \cite{JMMS}
provided feedback-form representations of MVH strategies for general model frameworks using semimartingale approaches, duality approaches or backward stochastic differential equations,
but their representations are not given concretely for concrete models.
Here, a representation of the MVH strategy $\tvtH_t$ is said to be feedback-form if it is given as $\tvtH_t=a_t+b_tG_{t-}(\tvtH)$ for some predictable processes $a$ and $b$.
Moreover, Lim \cite{Lim} considered a L\'evy market and gave a closed-form expression of $\tvtH$, that is, an expression which does not include any value of $\tvtH$ up to time $t-$.
However, he restricted $H$ to be bounded, and his expression is not an explicit form, since it includes solutions to backward stochastic differential equations.
On the other hand, as researches on explicit representations for concrete models,
Hubalek et al. \cite{HKK} obtained a representation in feedback-form for exponential L\'evy models,
and also their results have been extended to the additive process case and affine stochastic volatility models by Goutte et al. \cite{GOR}, and Kallsen and Vierthauer \cite{KV},
respectively.
The discussion in \cite{HKK} is based on bilateral Laplace transforms and the F\"ollmer-Schweizer (FS) decomposition,
which is an expression of $H$ by the sum of a stochastic integral with respect to $S$ and a residual martingale.
In addition, combining their Theorems 3.1 and 3.3, they also gave an explicit closed-form representation.

As mentioned before Theorem 3.3 in \cite{HKK}, a closed-form representation might be preferred to one in feedback-form from a numerical-analytical point of view,
since $G_{t-}(\tvtH)$ is approximated with an involved recursive calculation, which is very time-consuming and entails a drop of the accuracy.
For more details on this matter, see Remark \ref{rem-rec} in this paper.
In addition, the closed-form representation obtained in \cite{HKK} is also not appropriate to develop an efficient numerical scheme of $\tvtH_t$ for any given time $t\in[0,T]$
for the following two reasons:
First, their closed-form representation is given as a direct extension of the feedback-form one using a general stochastic exponential,
that is, an involved recursive calculation is still needed in order to compute $\tvtH_t$.
Second, a stochastic integral with respect to the quadratic variation of $S$ is included in their closed-form representation, but the quadratic variation of $S$ is not observable.
Therefore, in order to develop an efficient numerical method, we need to derive a new explicit closed-form representation which does not include a stochastic integration
with respect to unobservable data.
This is the first main purpose of this paper.
To this end, making use of results of \cite{AS}, we get a representation of $\tvtH$ by means of Malliavin calculus for L\'evy processes.
Furthermore, we rely on the argument of \cite{A05}, which is based on a different decomposition of $H$ from the FS decomposition.
As one more advantage of our representation, path-dependent options are covered as seen in Examples \ref{ex1} and \ref{ex3}, while \cite{HKK} excluded them.

Using the obtained closed-form representation of $\tvtH$, we shall develop an FFT-based numerical method to compute $\tvtH_t$ for any given $t\in[0,T]$ for exponential L\'evy models.
There is some literature on numerical analysis of MVH strategies for jump type models, say De Franco et al. \cite{DTW}, \cite{GOR}, \cite{HKK} and so on,
all of which computed the initial hedge $\tvtH_0$ or the hedging error induced by MVH strategies or both alone,
but no one has developed numerical schemes of $\tvtH_t$ for any given $t\in[0,T]$ so far as we know.
The most difficulty lies in the fact that $\tvtH_t$ is depending on the whole trajectory of $S$ up to time $t-$.
From a practical point of view, we cannot observe the trajectory of $S$ continuously, however.
Thus, we compute $\tvtH_t$ approximately using discrete observational data of $S$.
The numerical method developed in this paper has the following three features.
First, our method does not need any involved recursive calculation, although a simple recursive calculation is needed in order to compute a discretization of a stochastic exponential.
Second, we use discrete observational data of $S$ alone, in other words, our method do not need any unobservable data.
Third, we make use of results of Arai et al. \cite{AIS}, which has developed an FFT-based numerical scheme of LRM strategies for exponential L\'evy models using results of \cite{AS}.
In principle, computing an MVH strategy is time-consuming since its expression includes stochastic integrals.
Nevertheless, an FFT-based approach enables us to achieve high speed computation.

An outline of this paper is as follows:
Model description and mathematical preliminaries are given in Section 2.
In particular, we introduce the variance-optimal martingale measure and Malliavin calculus for L\'evy processes.
The main result and its proof are addressed in Section 3.
Subsection 3.2 introduces some examples.
Section 4 is devoted to a develop numerical scheme and to introduce numerical results.

%
%
\setcounter{equation}{0}
\section{Preliminaries}
\subsection{Model description}
We consider throughout a financial market being composed of one risky asset and one riskless asset with finite time horizon $T>0$.
For simplicity, we assume that the interest rate of the market is given by $0$, that is, the price of the riskless asset is $1$ at all times.

Let $(\Omega_W, \calF_W, \bbP_W)$ be a one-dimensional Wiener space on $[0,T]$; and $W$ its coordinate mapping process, that is, a one-dimensional standard Brownian motion with $W_0=0$.
$(\Omega_J, \calF_J, \bbP_J)$ denotes the canonical L\'evy space for a pure jump L\'evy process $J$ on $[0,T]$ with L\'evy measure $\nu$,
that is, $\Omega_J=\cup_{n=0}^\infty([0,T]\times\bbR_0)^n$,
where $J_t(\omega_J)=\sum_{i=1}^nz_i{\bf 1}_{\{t_i\leq t\}}$ for $t\in[0,T]$ and $\omega_J=((t_1,z_1),\dots,(t_n,z_n))\in([0,T]\times\bbR_0)^n$.
Note that $([0,T]\times\bbR_0)^0$ represents an empty sequence.
For more details on the canonical L\'evy space, see Sol\'e et al. \cite{S07}.
Now, we assume that $\int_{\bbR_0}z^2\nu(dz)<\infty$; and denote
\[
(\Omega, \calF, \bbP)=(\Omega_W\times\Omega_J, \calF_W\times\calF_J, \bbP_W\times\bbP_J).
\]
Let $\bbF=\{\calF_t\}_{t\in[0,T]}$ be the canonical filtration completed for $\bbP$.
Let $X$ be a square integrable centered L\'evy process on $(\Omega, \calF, \bbP)$ represented as
\[
X_t=W_t+J_t-t\int_{\bbR_0}z\nu(dz).
\]
Denoting by $N$ the Poisson random measure defined as $N(t,A):=\sum_{s\leq t}{\bf 1}_A(\Delta X_s)$, $A\in\calB(\bbR_0)$ and $t\in[0,T]$,
where $\Delta X_s:=X_s-X_{s-}$, we have $J_t=\int_0^t\int_{\bbR_0}zN(ds,dz)$.
In addition, we define its compensated measure as $\tN(dt,dz):=N(dt,dz)-\nu(dz)dt$.
Thus, $X$ is expressed as
\begin{equation}\label{eq-X}
X_t=W_t+\int_0^t\int_{\bbR_0}z\tN(du,dz).
\end{equation}

The fluctuation of the risky asset is assumed to be given by a solution to the following stochastic differential equation:
\begin{equation}\label{SDE}
dS_t=S_{t-}\l[\alpha_tdt+\beta_tdW_t+\int_{\bbR_0}\gamma_{t,z}\tN(dt,dz)\r],\hspace{3mm}S_0>0,
\end{equation}
where $\alpha$ and $\beta$ are deterministic measurable functions on $[0,T]$, and $\gamma$ is a deterministic jointly measurable function on $[0,T]\times\bbR_0$.
In addition, we denote
\[
\Gamma_t:=\int_{\bbR_0}\gamma^2_{t,z}\nu(dz), \ \ \ \hbox{and} \ \ \lambda_t:=\dfrac{\alpha_t}{S_{t-}(\beta_t^2+\Gamma_t)}
\]
for $t\in[0,T]$.
Now, we assume throughout this paper the following:

\begin{ass}\label{ass1}
\begin{enumerate}
\item $\gamma_{t,z}>-1$ for any $(t,z)\in[0,T]\times\bbR_0$.
\item $\sup_{t\in[0,T]}(|\alpha_t|+\beta^2_t+\Gamma_t)<C$ for some $C>0$.
\item There exists an $\ve>0$ such that
      \[
      \lambda_tS_{t-}\gamma_{t,z}<1-\ve \ \ \ \mbox{ and } \ \ \ \beta^2_t+\Gamma_t>\ve, \ (t,z,\omega)\mbox{-a.e.}
      \]
\end{enumerate}
\end{ass}

\begin{rem}\label{rem1}
\begin{enumerate}
\item Under Assumption \ref{ass1}, (\ref{SDE}) has a solution $S$ satisfying the so-called structure condition (SC), that is, $S$ has the following three properties:
      \begin{enumerate}
      \item $S$ is a semimartingale of the space $\calS^2$, that is, a special semimartingale with the canonical decomposition $S=S_0+M+A$ such that
            \begin{equation}\label{eq-S2}
            \l\|[M]_T^{1/2}+\int_0^T|dA_s|\r\|_{L^2(\bbP)}<\infty,
            \end{equation}
            where $dM_t=S_{t-}(\beta_tdW_t+\int_{\bbR_0}\gamma_{t,z}\tN(dt,dz))$ and $dA_t=S_{t-}\alpha_tdt$. \\
      \item We have $A=\int\lambda d\langle M\rangle$. \\
      \item The mean-variance trade-off process $\whK_t:=\int_0^t\lambda^2_sd\langle M\rangle_s$ is finite, that is, $\whK_T$ is finite $\bbP$-a.s. \\
      \end{enumerate}
      The SC is closely related to the no-arbitrage condition.
      For more details on the SC, see \cite{Sch} and \cite{Sch3}.
\item The process $\whK$ as well as $A$ is continuous. In particular, $\whK$ is deterministic.
\item (\ref{eq-S2}) implies that $\sup_{t\in[0,T]}|S_t|\in L^2(\bbP)$ by Theorem V.2 of Protter \cite{P}.
\item Item 1 in Assumption \ref{ass1} ensures the positivity of $S$.
      Thus, under Assumption \ref{ass1}, $S$ is an exponential of an additive process, that is, its logarithm $\log(S)$ has the following properties:
      \begin{enumerate}
      \item (Continuous in probability): $\log(S)$ has no fixed time of discontinuity.
      \item (Independent increments):    $\log(S_{t_1})-\log(S_{t_2})$ is independent of $\calF_{t_2}$ for $0\leq t_2< t_1\leq T$.
      \end{enumerate}
\end{enumerate}
\end{rem}

\subsection{Variance-optimal martingale measure}
In this subsection, we discuss the variance-optimal martingale measure, which is indispensable to discuss MVH strategies.
Roughly speaking, the variance-optimal martingale measure is defined as an equivalent martingale measure whose density minimizes its $L^2(\bbP)$-norm.

We start with the definition of admissible strategies.
We denote by $\Theta$ the space of all $\bbR$-valued predictable $S$-integrable processes $\vt$
whose stochastic integral $\int_0^t\vt_udS_u$ is a semimartingale of the space $\calS^2$.
Throughout this paper, we regard $\Theta$ as the set of all admissible strategies.
Remark that Assumption \ref{ass1} implies
\[
\Theta=\l\{\vt\mbox{ $\bbR$-valued predictable $S$-integrable process}: \bbE\l[\int_0^T\vt_u^2S_{u-}^2du\r]<\infty\r\}
\]

Next, we define the variance-optimal martingale measure as follows:

\begin{defn}[Section 1of Schweizer \cite{Sch0}]
\begin{enumerate}
\item A signed measure $\bbQ$ on $(\Omega, \calF)$ is called a signed $\Theta$-martingale measure if $\bbQ(\Omega)=1$, $\bbQ\ll\bbP$, $d\bbQ/d\bbP\in L^2(\bbP)$ and
      \[
      \bbE\l[\frac{d\bbQ}{d\bbP}\cdot G_T(\vt)\r]=0
      \]
      for any $\vt\in\Theta$, where $G_T(\vt)=\int_0^T\vt_udS_u$.
      We denote by $\calP_s(\Theta)$ the set of all signed $\Theta$-martingale measures, and $\calD_s(\Theta):=\{d\bbQ/d\bbP|\bbQ\in\calP_s(\Theta)\}$.
\item $\bbQ\in\calP_s(\Theta)$ is called an equivalent martingale measure if it is a probability measure equivalent to $\bbP$.
\item An equivalent martingale measure $\tP\in\calP_s(\Theta)$ is called the variance-optimal martingale measure if its density $d\tP/d\bbP$ minimizes $\|D\|_{L^2(\bbP)}$ over all $D\in\calD_s(\Theta)$.
\end{enumerate}
\end{defn}

We shall show that the variance-optimal martingale measure is given by a probability measure $\tP$ defined as
\[
\frac{d\tP}{d\bbP}=\calE_T\l(-\int_0^\cdot\lambda_udM_u\r)
\]
where $\calE(Y)$ represents the stochastic exponential of $Y$, that is, a solution to the stochastic differential equation $d\calE_t(Y)=\calE_{t-}(Y)dY_t$ with $\calE_0(Y)=1$.
Henceforth, we denote $D^*:=\frac{d\tP}{d\bbP}$ and $Z_t:=\calE_t(-\int_0^\cdot\lambda_udM_u)$.
Note that $Z$ is a positive square integrable martingale under Assumption \ref{ass1} as seen in Example 2.8 of \cite{AS}, that is, $\tP$ is an equivalent martingale measure.
We prove the following proposition in order to make sure that our setting satisfies all the standing assumptions in \cite{A05}.

\begin{prop}\label{prop1}
Under Assumption \ref{ass1}, we have the following:
\begin{enumerate}
\item $\tP$ is the variance-optimal martingale measure.
\item $Z$ satisfies the reverse H\"older inequality $R_2(\bbP)$, that is, there is a constant $C>0$ such that, for any stopping time $\tau\leq T$, we have
      \[
      \bbE\l[\l(\frac{Z_T}{Z_\tau}\r)^2\Big|\calF_\tau\r]\leq C.
      \]
\item There is a $C>0$ such that $Z_{t-}\leq CZ_t$ for any $t\in[0,T]$.
\end{enumerate}
\end{prop}

\proof
To see item 1, we define $\tZ_t:=\bbE_{\tP}[D^*|\calF_t]$.
Note that $\tZ_T=D^*$ and $\tZ_0=\bbE[(D^*)^2]$.
Now, we calculate $\tZ$ as follows:
\begin{align*}
\tZ_t&= \bbE_{\tP}\l[\calE_T\l(-\int_0^\cdot\lambda_udM_u\r)\Big|\calF_t\r] \\
     &= \exp\l(\int_0^T\lambda_udA_u\r)\bbE_{\tP}\l[\calE_T\l(-\int_0^\cdot\lambda_udS_u\r)\Big|\calF_t\r] \\
     &= \exp\l(\int_0^T\lambda_udA_u\r)\calE_t\l(-\int_0^\cdot\lambda_udS_u\r)=\bbE[(D^*)^2]\calE_t\l(-\int_0^\cdot\lambda_udS_u\r)
\end{align*}
for any $t\in[0,T]$, since $\int_0^\cdot\lambda_udA_u$ is deterministic by Remark \ref{rem1}.
Note that $\tZ$ is a solution to the following equation:
\[
\tZ_t=\tZ_0-\int_0^t\tZ_{u-}\lambda_udS_u.
\]
For any $\bbQ\in\calP_s(\Theta)$, we have
\begin{align*}
\l\|\frac{d\bbQ}{d\bbP}\r\|^2_{L^2(\bbP)}
&=    \l\|\frac{d\bbQ}{d\bbP}-D^*\r\|^2_{L^2(\bbP)}+2\bbE\l[\frac{d\bbQ}{d\bbP}D^*\r]-\|D^*\|^2_{L^2(\bbP)} \\
&\geq 2\tZ_0\bbE\l[\frac{d\bbQ}{d\bbP}\calE_T\l(-\int_0^\cdot\lambda_udS_u\r)\r]-\tZ_0=\tZ_0,
\end{align*}
from which item 1 follows.

Item 2 holds true from Proposition 3.7 of Choulli et al. \cite{CKS}.
In addition, since we can find an $\ve>0$ such that $\lambda_t\Delta M_t<1-\ve$ for any $t\in[0,T]$ due to item 3 of Assumption \ref{ass1}, we have
$Z_t/Z_{t-}=1-\lambda_t\Delta M_t>1-(1-\ve)=\ve$, from which item 3 follows.
\fin

\begin{rem}
The essence of the above proof lies in the fact that $\int_0^T\lambda_udA_u$($=\whK_T$) is deterministic.
General speaking, when $\whK_T$ is deterministic, the variance-optimal martingale measure coincides with the minimal martingale measure,
which is defined as an equivalent martingale measure under which any square-integrable $\bbP$-martingale orthogonal to $M$ remains a martingale.
Actually, Example 2.8 of \cite{AS} showed that $\tP$ is the minimal martingale measure.
Note that the minimal martingale measure is essential in the LRM approach.
\end{rem}

Now, we prepare some notations for later use.
The Girsanov theorem implies that
\[
W^{\tP}_t:=W_t+\int_0^t\lambda_uS_{u-}\beta_udu
\]
is a one-dimensional standard Brownian motion under $\tP$.
Moreover, denoting
\begin{equation}\label{eq-nu*}
\nu^{\tP}_t(dz):=(1-\lambda_tS_{t-}\gamma_{t,z})\nu(dz),
\end{equation}
and $\tN^{\tP}(dt,dz):=N(dt,dz)-\nu^{\tP}_t(dz)dt$ for $t\in[0,T]$ and $z\in\bbR_0$, we have
\[
\bbE_{\tP}\l[\tN^{\tP}(A,B)\r]=0
\]
for any $A\in\calB([0,T])$ and $B\in\calB(\bbR_0)$.
Hence, we can rewrite the stochastic differential equation (\ref{SDE}) as
\begin{equation}\label{SDE2}
dS_t=S_{t-}\l[\beta_tdW^{\tP}_t+\int_{\bbR_0}\gamma_{t,z}\tN^{\tP}(dt,dz)\r],\hspace{3mm}S_0>0.
\end{equation}

\subsection{Malliavin calculus}
One of our aims in this paper is to obtain a closed-from representation of MVH strategies in terms of Malliavin calculus for L\'evy processes.
Now, we introduce some notations and definitions related to the Malliavin calculus.
We adapt the canonical L\'evy space framework undertaken by \cite{S07}, which is a Malliavin calculus on the L\'evy process $X$ given in (\ref{eq-X}).
First of all, we define measures $q$ and $Q$ on $[0,T]\times\bbR$ as
\[
q(E):=\int_E\delta_0(dz)dt+\int_Ez^2\nu(dz)dt,
\]
and
\[
Q(E):=\int_E\delta_0(dz)dW_t+\int_Ez\tN(dt,dz),
\]
where $E\in\calB([0,T]\times\bbR)$ and $\delta_0$ is the Dirac measure at $0$.
For $n\in\bbN$, we denote by $L_{T,q,n}^2$ the set of product measurable, deterministic functions $f:([0,T]\times\bbR)^n\to\bbR$ satisfying
\[
\|f\|_{ L_{T,q,n}^2}^2:=\int_{([0,T]\times\bbR)^n}|f((t_1,z_1),\cdots,(t_n,z_n))|^2q(dt_1,dz_1)\cdots q(dt_n,dz_n)<\infty.
\]
For $n\in\bbN$ and $f\in L_{T,q,n}^2$, we define
\[
I_n(f):=\int_{([0, T]\times\bbR)^n}f((t_1,z_1),\cdots,(t_n,z_n))Q(dt_1,dz_1)\cdots Q(dt_n,dz_n).
\]
Formally, we denote $L_{T,q,0}^2:=\bbR$ and $I_0(f):=f$ for $f\in\bbR$.
Under this setting, any $F\in L^2(\bbP)$ has the unique representation $F=\sum_{n=0}^{\infty}I_n(f_n)$ with functions $f_n\in L_{T,q,n}^2$ that are symmetric
in the $n$ pairs $(t_i,z_i), 1\leq i\leq n$, and we have $\bbE[F^2]=\sum_{n=0}^{\infty }n!\|f_n\|_{L_{T,q,n}^2}^2$.
Now, we define a Malliavin derivative operator $D$ as follows:

\begin{defn}
\begin{enumerate}
\item Let $\bbD^{1,2}$ denote the set of random variables $F\in L^2(\bbP)$ with $F=\sum_{n=0}^\infty I_n(f_n)$ satisfying $\sum_{n=1}^\infty nn!\|f_n\|_{L_{T,q,n}^2}^2<\infty$.
\item For any $F\in\bbD^{1,2}$, $DF:[0,T]\times\bbR\times\Omega\to\bbR$ is defined by
\[
D_{t,z}F=\sum_{n=1}^\infty nI_{n-1}(f_n((t,z),\cdot))
\]
for $q$-a.e. $(t,z)\in[0,T]\times\bbR$, $\bbP$-a.s.
\end{enumerate}
\end{defn}

Let $H\in L^2(\bbP)$ be a random variable representing the payoff of a claim to hedge.
In addition to Assumption \ref{ass1}, we assume throughout the following:

\begin{ass}\label{ass2}
\begin{enumerate}
\item $Z_TH\in L^2(\bbP)$.
\item $H\in\bbD^{1,2}$ and $Z_TD_{t,z}H+HD_{t,z}Z_T+zD_{t,z}H\cdot D_{t,z}Z_T\in L^2(q\times\bbP)$.
\end{enumerate}
\end{ass}

\noindent
Note that Assumption \ref{ass2} is not restrictive for $H$.
Indeed, many of typical claims satisfy Assumption \ref{ass2} under mild conditions as seen in subsection 3.2.
Under Assumptions \ref{ass1} and \ref{ass2}, Example 3.9 of \cite{AS} implies that $H$ is described as
\begin{equation}\label{eq-CO}
H=\bbE_{\tP}[H]+\int_0^TI_tdW^{\tP}_t+\int_0^T\int_{\bbR_0}J_{t,z}\tN^{\tP}(dt,dz),
\end{equation}
where $I_t:=\bbE_{\tP}[D_{t,0}H|\calF_{t-}]$, and $J_{t,z}:=\bbE_{\tP}[zD_{t,z}H|\calF_{t-}]$ for $t\in[0,T]$ and $z\in\bbR_0$.
Now, we denote additionally for later use $K_t:=\int_{\bbR_0}J_{t,z}\gamma_{t,z}\nu(dz)$ and
\[
H_t:=\bbE_{\tP}[H|\calF_t]=\bbE_{\tP}[H]+\int_0^tI_udW^{\tP}_u+\int_0^t\int_{\bbR_0}J_{u,z}\tN^{\tP}(du,dz)
\]
for $t\in[0,T]$.

%
%
\setcounter{equation}{0}
\section{Main results}
We derive in this section a closed-from expression of the MVH strategy for a claim $H\in L^2(\bbP)$ in terms of Malliavin derivatives.
As mentioned in Introduction, the MVH strategy for $H$ is defined as a pair $(\tcH,\tvtH)\in\bbR\times\Theta$ which minimizes
\begin{equation}\label{eq-MVH}
\min_{c\in\bbR, \vt\in\Theta}\bbE\l[\l(H-c-G_T(\vt)\r)^2\r].
\end{equation}
The following theorem is shown in Subsection 3.1, and some examples will be introduced in Subsection 3.2.

\begin{thm}\label{main-thm}
Under Assumptions \ref{ass1} and \ref{ass2}, the MVH strategy $(\tcH,\tvtH)\in\bbR\times\Theta$ for claim $H\in L^2(\bbP)$ is represented in closed-form as
\[
\tcH=\bbE_{\tP}[H]
\]
and
\begin{equation}\label{eq-main-thm}
\tvtH_t=\txiH_t+\lambda_t\calE_{t-}\bigg(\int_0^{t-}\frac{dH_u-\txiH_udS_u}{\calE_u}
        +\int_0^{t-}\frac{\alpha_u\beta_u(\Gamma_uI_u-\beta_uK_u)}{\calE_{u-}(\beta_u^2+\Gamma_u)^2}du\bigg)
\end{equation}
for $t\in[0,T]$, where $\calE_t:=\tZ_t/\tZ_0=\calE_t(-\int_0^\cdot\lambda_udS_u)$ and
\begin{equation}\label{eq-LRM}
\txiH_t:=\frac{\beta_t I_t+K_t}{S_{t-}(\beta^2_t+\Gamma_t)}.
\end{equation}
\end{thm}

\begin{rem}
As shown in \cite{AS}, $\txiH$ defined in (\ref{eq-LRM}) represents the LRM strategy for claim $H$, that is,
for each $t\in[0,T]$, $\txiH_t$ gives the number of shares of the risky asset in the LRM strategy at time $t$.
\end{rem}

\begin{rem}
Benth et al. \cite{B} treated MVH strategies using Malliavin calculus for L\'evy processes under the assumption that $S$ is a martingale.
This assumption is very restrictive, and simplifies the problem.
\end{rem}

\begin{ex}[Exponential L\'evy models]\label{ex2}
As a typical and simple model framework of $S$, we consider exponential L\'evy models, that is, the case where $\log(S_t/S_0)$ is a L\'evy process represented as
\[
\log(S_t)-\log(S_0)=\mu t+\sigma W_t+\int_{\bbR_0}z\tN([0,t],dz)
\]
for $t\in[0,T]$, where $\mu\in\bbR$, $\sigma\geq0$.
Under Assumption \ref{ass1}, $S$ is a solution to the following stochastic differential equation:
\[
dS_t=S_{t-}\l(\mu^Sdt+\sigma dW_t+\int_{\bbR_0}(e^z-1)\tN(dt,dz)\r),
\]
where $\mu^S=\mu+\frac{1}{2}\sigma^2 +\int_{\bbR_0}(e^z-1-z)\nu(dz)$.
Supposing Assumption \ref{ass2} additionally, we have, by Theorem \ref{main-thm}
\begin{align}
\tvtH_t=\txiH_t+\frac{\mu^S\calE_{t-}}{S_{u-}(\sigma^2+\Gamma)}\bigg(\int_0^{t-}\frac{dH_u-\txiH_udS_u}{\calE_u}
        +\int_0^{t-}\frac{\mu^S\sigma(\Gamma I_u-\sigma K_u)}{\calE_{u-}(\sigma^2+\Gamma)^2}du\bigg),
\label{eq-exp-Levy}
\end{align}
where $\Gamma:=\int_{\bbR_0}(e^z-1)^2\nu(dz)$.
As seen in \cite{AS}, the condition
\[
\int_{\bbR_0}\l\{z^2+(e^z-1)^4\r\}\nu(dz)<\infty
\]
guarantees Assumption \ref{ass2} for options introduced in Subsection 3.2.
In addition, \cite{AIS} introduced a numerical method to compute $I_t$, $K_t$, $H_t$ and $\txiH_t$ for the case where $H$ is a call option.
\end{ex}

\subsection{Proof of Theorem \ref{main-thm}}
{\it Step 1:} \ \ \
\cite{A05} obtained a similar feedback-form representation to \cite{HKK} for more general discontinuous semimartingale models.
In \cite{A05}, he defined a new decomposition of $H$, which is different from the FS decomposition.
Remark that \cite{A05} treated, instead of (\ref{eq-MVH}), the following minimization problem:
\begin{equation}\label{eq-MVH2}
\min_{\vt\in\Theta}\bbE\l[\l(H-G_T(\vt)\r)^2\r].
\end{equation}
Now, we introduce an outline of the argument in \cite{A05} as a preparation step.

Recall that Proposition \ref{prop1} holds true under Assumption \ref{ass1}.
This fact ensures that our setting satisfies Assumption 1 of \cite{A05}.
Thus, the solution to (\ref{eq-MVH2}) exists, and we denote it by $\hvtH\in\Theta$.
We define a new probability measure $\tQ$ as
\[
\frac{d\tQ}{d\tP}=\frac{\tZ_T}{\tZ_0}=\calE_T\l(-\int_0^\cdot\lambda_udS_u\r).
\]
As shown in (4.7) of \cite{A05}, $H$ admits the following decomposition:
\begin{equation}\label{eq-new}
H=\bbE_{\tP}[H]+G_T(\hetaH)+\hNH_T,
\end{equation}
where $\hetaH\in\Theta$, and $\hNH$ is a $\tP$-martingale with $\hNH_0=0$.
Here $\hNH$ is represented as
\begin{equation}\label{eq-defNH}
\hNH_t=\int_0^t\tZ_{u-}d\hLH_u+[\tZ,\hLH]_t
\end{equation}
with a square integrable $\tQ$-martingale $\hLH$.
Note that the processes $\hLH$ and $S\hLH$ both are $\tP$-martingales.
Remark that the decomposition (\ref{eq-new}) is neither the Kunita-Watanabe one nor the FS one in our setting.
Furthermore, (4.5) in \cite{A05} provides that $\hvtH$ is given by
\begin{equation}\label{eq-hvtH}
\hvtH_t=\hetaH_t+\bbE_{\tP}[H]\lambda_t\calE_{t-}+\hLH_{t-}\lambda_t\tZ_{t-}.
\end{equation}

{\it Step 2:} \ \ \
Replacing $H$ with the constant $1$ in (\ref{eq-MVH2}), we consider the following minimization problem:
\begin{equation}\label{proj1}
\min_{\vt\in\Theta}\bbE\l[\l(1-G_T(\vt)\r)^2\r].
\end{equation}
Letting $\hvt^1_t:=\lambda_t\calE_{t-}$, we have $\hvt^1\in\Theta$, since $\bbE[\int_0^T(\hvt^1)_u^2S_{u-}^2du]\leq C\bbE[Z_T^2]<\infty$ for some $C>0$.
In addition, we have $1-G_T(\hvt^1)=\calE_T$,
\[
\bbE[\calE_TG_T(\vt)]=\tZ_0^{-1}\bbE[\tZ_TG_T(\vt)]=\tZ_0^{-1}\bbE_{\tP}[G_T(\vt)]=0
\]
for any $\vt\in\Theta$, and $\bbE[\calE_T^2]=\bbE[\calE_T(1-G_T(\hvt^1))]=\bbE[\calE_T]$.
Thus, we obtain
\begin{align*}
\lefteqn{\bbE[(1-G_T(\vt))^2]} \\
&=    \bbE[(G_T(\vt)-G_T(\hvt^1))^2]+2\bbE[\calE_T(1-G_T(\vt))]-\bbE[(1-G_T(\hvt^1))^2] \\
&\geq 2\bbE[\calE_T]-\bbE[(1-G_T(\hvt^1))^2]=\bbE[(1-G_T(\hvt^1))^2]
\end{align*}
for any $\vt\in\Theta$, which means that $\hvt^1$ is the solution to (\ref{proj1}).
Hence, Theorem 4.2 of Hou and Karatzas \cite{HK} implies that $\tcH=\bbE_{\tP}[H]$, and $\hvtHc=\hvtH-c\hvt^1$ for any $c\in\bbR$,
where $\hvtHc$ is the solution to the following minimization problem:
\[
\min_{\vt\in\Theta}\bbE\l[\l(H-c-G_T(\vt)\r)^2\r]
\]
for fixed $c\in\bbR$.
As a result, we obtain from (\ref{eq-hvtH})
\begin{equation}\label{eq-tvtH}
\tvtH_t=\hvt^{H-\tcH}=\hvtH-\bbE_{\tP}[H]\hvt^1=\hetaH_t+\hLH_{t-}\lambda_t\tZ_{t-}.
\end{equation}

{\it Step 3:} \ \ \
All we have to do is to derive representations of $\hetaH$ and $\hLH$.
To this end, we prepare some notations.
The Girsanov theorem implies that
\[
\l\{
\begin{array}{l}
W^{\tQ}_t:=W^{\tP}_t+\int_0^t\lambda_uS_{u-}\beta_udu \\
\tN^{\tQ}(dt,dz):=\tN^{\tP}(dt,dz)+\lambda_tS_{t-}\gamma_{t,z}\nu_t^{\tP}(dz)dt \\
\end{array}\r.
\]
are a one-dimensional Brownian motion under $\tQ$, and the compensated Poisson random measure of $N$ under $\tQ$, respectively.
From the view of the martingale representation property under $\tQ$ (Theorem III.4.34 of Jacod and Shiryaev \cite{JS03}), $\hLH$ is described as, for any $t\in[0,T]$,
\begin{equation}\label{eq-hLH}
\hLH_t=\int_0^t\hlW_udW^{\tQ}_u+\int_0^t\int_{\bbR_0}\hlN_{u,z}\tN^{\tQ}(du,dz)
\end{equation}
for some predictable processes $\hlW$ and $\hlN$.
Since the product process $S\hLH$ is a martingale under $\tP$, we get
\begin{equation}\label{eq-ortho}
\beta_t\hlW_t+\int_{\bbR_0}\gamma_{t,z}\hlN_{t,z}\nu_t^{\tP}(dz)=0
\end{equation}
for any $t\in[0,T]$.
Thus, we  can rewrite (\ref{eq-hLH}) as
\begin{equation}\label{eq-hLH3}
\hLH_t=\int_0^t\hlW_udW^{\tP}_u+\int_0^t\int_{\bbR_0}\hlN_{u,z}\tN^{\tP}(du,dz).
\end{equation}

Next, (\ref{eq-defNH}) provides
\begin{equation}\label{eq-hNH}
\hNH_T=\int_0^T\tZ_{u-}\hlW_udW^{\tP}_u+\int_0^T\int_{\bbR_0}\tZ_{u-}\hlN_{u,z}(1-\lambda_uS_{u-}\gamma_{u,z})\tN^{\tP}(du,dz),
\end{equation}
since the condition (\ref{eq-ortho}), together with (\ref{SDE2}), implies that
\begin{align*}
[\tZ,\hLH]_T &= -\int_0^T\lambda_u\tZ_{u-}d[S,\hLH]_u \\
             &= -\int_0^T\lambda_u\tZ_{u-}S_{u-}\l(\beta_u\hlW_udu+\int_{\bbR_0}\gamma_{u,z}\hlN_{u,z}N(du,dz)\r) \\
             &= -\int_0^T\lambda_u\tZ_{u-}S_{u-}\int_{\bbR_0}\gamma_{u,z}\hlN_{u,z}\tN^{\tP}(du,dz).
\end{align*}
Then, (\ref{eq-new}) together with (\ref{SDE2}) and (\ref{eq-hNH}) provides
\begin{align}
H&= \bbE_{\tP}[H]+\int_0^T(\hetaH_uS_{u-}\beta_u+\tZ_{u-}\hlW_u)dW^{\tP}_u \nonumber \\
 &  \hspace{5mm}+\int_0^T\int_{\bbR_0}\l(\hetaH_uS_{u-}\gamma_{u,z}+\tZ_{u-}\hlN_{u,z}(1-\lambda_uS_{u-}\gamma_{u,z})\r)\tN^{\tP}(du,dz).
\label{eq-H}
\end{align}

Comparing (\ref{eq-H}) with (\ref{eq-CO}), we obtain, for any $t\in[0,T]$,
\begin{equation}\label{eq-sim}
\l\{
\begin{array}{l}
   \hetaH_tS_{t-}\beta_t+\tZ_{t-}\hlW_t=I_t, \\
   \hetaH_tS_{t-}\gamma_{t,z}+\tZ_{t-}\hlN_{t,z}(1-\lambda_tS_{t-}\gamma_{t,z})=J_{t,z},
\end{array}\r.
\end{equation}
which yields
\[
\hlW_t=\frac{1}{\tZ_{t-}}(I_t-\hetaH_tS_{t-}\beta_t),
\]
and
\[
\hlN_{t,z}(1-\lambda_tS_{t-}\gamma_{t,z})=\frac{1}{\tZ_{t-}}\l(J_{t,z}-\hetaH_tS_{t-}\gamma_{t,z}\r).
\]
Solving the simultaneous equation (\ref{eq-sim}) on $\hetaH$ by using (\ref{eq-nu*}) and (\ref{eq-ortho}), we have
\[
\hetaH_t=\frac{\beta_tI_t+K_t}{S_{t-}(\beta_t^2+\Gamma_t)}.
\]
(\ref{eq-LRM}) implies that $\hetaH$ coincides with $\txiH$, which is the LRM strategy for $H$.

Consequently, we get
\[
\hlW_t=\frac{1}{\tZ_{t-}}\frac{\Gamma_tI_t-\beta_tK_t}{\beta_t^2+\Gamma_t}
\]
as well as
\[
\hlN_{t,z}(1-\lambda_tS_{t-}\gamma_{t,z})=\frac{1}{\tZ_{t-}}\l(J_{t,z}-\frac{\beta_tI_t+K_t}{\beta_t^2+\Gamma_t}\gamma_{t,z}\r).
\]
Thus, (\ref{eq-hLH3}) implies that
\begin{equation}\label{eq-hLH2}
d\hLH_t=\frac{1}{\tZ_{t-}}\frac{\Gamma_tI_t-\beta_tK_t}{\beta_t^2+\Gamma_t}dW^{\tP}_t
        +\frac{1}{\tZ_{t-}}\int_{\bbR_0}\frac{J_{t,z}-\frac{\beta_tI_t+K_t}{\beta_t^2+\Gamma_t}\gamma_{t,z}}{1-\lambda_tS_{t-}\gamma_{t,z}}\tN^{\tP}(dt,dz).
\end{equation}

{\it Step 4:} \ \ \
From the view of (\ref{eq-tvtH}) together with (\ref{eq-hLH2}), we calculate $\tvtH$ as follows:
\begin{align*}
\tvtH_t
&=  \hetaH_t+\hLH_{t-}\lambda_t\tZ_{t-} \\
&=  \txiH_t+\lambda_t\tZ_{t-}\bigg(\int_0^{t-}\frac{1}{\tZ_{u-}}\frac{\Gamma_uI_u-\beta_uK_u}{\beta_u^2+\Gamma_u}dW^{\tP}_u \\
&   \hspace{5mm}+\int_0^{t-}\int_{\bbR_0}\frac{1}{\tZ_{u-}}\frac{1}{1-\lambda_uS_{u-}\gamma_{u,z}}\l(J_{u,z}-\frac{\beta_uI_u+K_u}{\beta_u^2+\Gamma_u}\gamma_{u,z}\r)\tN^{\tP}(du,dz)\bigg)
    \\
&=: \txiH_t+\lambda_t\tZ_{t-}(\Xi^W_{t-}+\Xi^N_{t-}).
\end{align*}
We have
\begin{align}
\Xi^W_{t-}
&= \int_0^{t-}\frac{1}{\tZ_{u-}}\frac{\Gamma_uI_u-\beta_uK_u}{\beta_u^2+\Gamma_u}dW^{\tP}_u \nonumber \\
&= \int_0^{t-}\frac{I_u}{\tZ_{u-}}dW^{\tP}_u-\int_0^{t-}\frac{\beta_uI_u+K_u}{\beta_u^2+\Gamma_u}\frac{\beta_u}{\tZ_{u-}}dW^{\tP}_u \nonumber \\
&= \int_0^{t-}\frac{I_u}{\tZ_u}dW^{\tP}_u-\int_0^{t-}\frac{\txiH_u}{\tZ_u}S_{u-}\beta_udW^{\tP}_u,
\label{eq-XiW}
\end{align}
and
\begin{align}
\Xi^N_{t-}
&= \int_0^{t-}\int_{\bbR_0}\frac{1}{\tZ_{u-}}\frac{1}{1-\lambda_uS_{u-}\gamma_{u,z}}\l(J_{u,z}-\frac{\beta_uI_u+K_u}{\beta_u^2+\Gamma_u}\gamma_{u,z}\r)\tN^{\tP}(du,dz) \nonumber \\
&= \int_0^{t-}\int_{\bbR_0}\frac{1}{\tZ_{u-}}\frac{1}{1-\lambda_uS_{u-}\gamma_{u,z}}(J_{u,z}-\txiH_uS_{u-}\gamma_{u,z})\tN^{\tP}(du,dz) \nonumber \\
&= \int_0^{t-}\int_{\bbR_0}(J_{u,z}-\txiH_uS_{u-}\gamma_{u,z})\l(\frac{N(du,dz)}{\tZ_u}-\frac{\nu_u^{\tP}(dz)du}{\tZ_{u-}(1-\lambda_uS_{u-}\gamma_{u,z})}\r) \nonumber \\
&= \int_0^{t-}\int_{\bbR_0}(J_{u,z}-\txiH_uS_{u-}\gamma_{u,z})\l(\frac{\tN^{\tP}(du,dz)}{\tZ_u}-\frac{\lambda_uS_{u-}\gamma_{u,z}\nu(dz)du}{\tZ_{u-}}\r) \nonumber \\
&= \int_0^{t-}\int_{\bbR_0}(J_{u,z}-\txiH_uS_{u-}\gamma_{u,z})\frac{\tN^{\tP}(du,dz)}{\tZ_u}+\int_0^{t-}\frac{\alpha_u\beta_u(\Gamma_uI_u-\beta_uK_u)}{\tZ_{u-}(\beta_u^2+\Gamma_u)^2}du,
\label{eq-XiN}
\end{align}
where the third equality is given from
\[
\int_{\bbR_0}\frac{1}{\tZ_{u-}}\frac{1}{1-\lambda_uS_{u-}\gamma_{u,z}}N(du,dz) = \int_{\bbR_0}\frac{N(du,dz)}{\tZ_u}.
\]
Thus, from (\ref{eq-XiW}) and (\ref{eq-XiN}), we obtain
\[
\Xi^W_{t-}+\Xi^N_{t-}=\int_0^{t-}\frac{dH_u-\txiH_udS_u}{\tZ_u}+\int_0^{t-}\frac{\alpha_u\beta_u(\Gamma_uI_u-\beta_uK_u)}{\tZ_{u-}(\beta_u^2+\Gamma_u)^2}du.
\]
This completes the proof of Theorem \ref{main-thm}.
\fin

\subsection{Examples}
\begin{ex}[Call and Asian options]\label{ex1}
We consider two representative options as contingent claims to hedge: call options $(S_T-K)^+$ and Asian options $(\frac{1}{T}\int_0^TS_udu-K)^+$ for $K>0$.
In order to obtain explicit representations of $\tvtH$ for such options, we have only to show expressions of $I_t$ and $J_{t,z}$ from the view of (\ref{eq-main-thm}).
In addition to Assumption \ref{ass1}, we assume the following condition:
\[
\int_{\bbR_0}\{\gamma_{t,z}^4+|\log(1+\gamma_{t,z})|^2\}\nu(dz)<C \ \mbox{ for some }C>0,
\]
which ensures Assumption \ref{ass2} as seen in Sections 4 and 5 of \cite{AS}.
For $K>0$ and $t\in[0,T]$, we have that
\[
\l\{
\begin{array}{lll}
   I_t    & = & \beta_t\bbE_{\tP}[{\bf 1}_{\{S_T>K\}}S_T|\calF_{t-}], \\
   J_{t,z}& = & \bbE_{\tP}[(S_T(1+\gamma_{t,z})-K)^+-(S_T-K)^+|\calF_{t-}]
\end{array}\r.
\]
for call options $(S_T-K)^+$, and
\[
\l\{
\begin{array}{lll}
I_t     & = & \beta_t\bbE_{\tP}[{\bf 1}_{\{V_0>K\}}V_t|\calF_{t-}], \\
J_{t,z} & = & \bbE_{\tP}\l[(V_0+\gamma_{t,z}V_t-K)^+-(V_0-K)^+|\calF_{t-}\r],
\end{array}\r.
\]
where $V_t=\frac{1}{T}\int_t^TS_udu$, for Asian options $(\frac{1}{T}\int_0^TS_udu-K)^+$.
\end{ex}

\begin{ex}[Lookback options]\label{ex3}
The payoff of a call option is given as a function of $S_T$.
On the other hand, that of an Asian option depends on the whole trajectory of $S$.
Such options are said to be path-dependent.
Various types of path-dependent options have been traded actively in recent years, but they are excluded in \cite{HKK}.
As one more typical example of path-dependent options covered by Theorem \ref{main-thm}, we deal with a lookback option, whose payoff depends on the running maximum of $S$.
In particular, we consider the case of $H=(M^S-K)^+$, where $M^S:=\sup_{t\in[0,T]}S_t$ and $K>0$.
For an exponential L\'evy model introduced in Example \ref{ex2}, Section 6 of \cite{AS} implies that,
under Assumption \ref{ass1} and the condition $\int_{\bbR_0}\l\{z^2+(e^z-1)^4\r\}\nu(dz)<\infty$,
\[
\l\{
\begin{array}{lll}
I_t     & = & \sigma\bbE_{\tP}[M^S{\bf 1}_{\{\log(M^S)>\log(K/S_0)\}}{\bf 1}_{\{\tau\geq t\}}|\calF_{t-}], \\
J_{t,z} & = & \bbE_{\tP}\l[\l(\sup_{u\in[0,T]}\l(S_ue^{z{\bf 1}_{\{t\leq u\}}}\r)-K\r)^+-(M^S-K)^+\Big|\calF_{t-}\r],
\end{array}\r.
\]
where $\tau:=\inf\{t\in[0,T]|S_t\vee S_{t-}=M^S\}$.
\end{ex}

%
%
\setcounter{equation}{0}
\section{Numerical analysis}
We shall develop a simple numerical scheme of the values of MVH strategy $\tvtH_t$ for any given time $t\in[0,T]$ for exponential L\'evy models,
and illustrate in subsection 4.1 some numerical results.
To our best knowledge, no numerical methods for the values of $\tvtH_t$ have been developed.
Remark that the value of $\tvtH_t$ is depending on not only $S_{t-}$ but also the whole trajectory of $S$ from $0$ to $t-$.
However, it is impossible to observe the trajectory of $S$ continuously from a practical point of view.
Accordingly, we compute $\tvtH_t$ approximately using discrete observational data $S_0,S_{t_1},S_{t_2},\dots$.

We consider in this section an exponential L\'evy model introduced in Example \ref{ex2}; and divide the time interval $[0,t]$ equally into
$0=t_0<t_1<\dots<t_{n+1}=t$ for $n\geq1$ for sake of simplicity.
Denote
\[
\l\{
\begin{array}{lll}
H_{t_k}     & = & \bbE_{\tP}[H|S_{t_k}], \\
I_{t_k}     & = & \bbE_{\tP}[D_{t_k,0}H|S_{t_{k-1}}],\vspace{0.5mm} \\
K_{t_k}     & = & \int_{\bbR_0}\bbE_{\tP}[zD_{t_k,z}H|S_{t_{k-1}}](e^z-1)\nu(dz),\vspace{1mm} \\
\txiH_{t_k} & = & \frac{\sigma I_{t_k}+K_{t_k}}{S_{t_{k-1}}(\sigma^2+\Gamma)}
\end{array}\r.
\]
for $k=1,\dots,n+1$.
Remark that the value of the MVH strategy at the initial date is given by $\tvtH_{t_1}$;
and all $H_{t_k}$, $I_{t_k}$ and $K_{t_k}$ are computable by the FFT-based approach developed in \cite{AIS}.
$\tvtH_t$ is then approximated from the view of (\ref{eq-exp-Levy}) as
\begin{equation}\label{approx}
\tvtH_t \approx \txiH_t+\frac{\mu^S\calE_{t_n}}{S_{t_n}(\sigma^2+\Gamma)}\bigg(\sum_{k=1}^n\frac{\Delta H_{t_k}-\txiH_{t_k}\Delta S_{t_k}}{\calE_{t_k}}
                +\sum_{k=1}^n\frac{\mu^S\sigma(\Gamma I_{t_k}-\sigma K_{t_k})}{\calE_{t_{k-1}}(\sigma^2+\Gamma)^2}\Delta t_k\bigg),
\end{equation}
where $\Delta X_{t_k}:=X_{t_k}-X_{t_{k-1}}$ for any sequence $\{X_{t_k}\}$, and $H_{t_0}:=\bbE_{\tP}[H]$.
Moreover, we approximate $\calE_{t_k}$ using a recursive calculation as $\calE_0=1$ and
\begin{equation}\label{eq-calE}
\calE_{t_{k+1}}=\calE_{t_k}\l\{1-\frac{\mu^S}{\sigma^2+\Gamma}\frac{\Delta S_{t_{k+1}}}{S_{t_k}}\r\}
               =\prod_{l=1}^k\l\{1-\frac{\mu^S}{\sigma^2+\Gamma}\frac{\Delta S_{t_{l+1}}}{S_{t_l}}\r\}
\end{equation}
for $k=1\dots,n$, which is a discretization of a stochastic exponential.

\begin{rem}\label{rem-rec}
An approximation of $\tvtH$ using a feedback-form expression is basically given as a discretization of a general stochastic exponential,
defined as a solution $Y$ to the following type of stochastic differential equation:
\[
Y_t=U_t+\int_0^tY_{u-}dV_u,
\]
where $U$ and $V$ are semimartingales.
Theorem V.52 of \cite{P} implies that, if $V$ is continuous, then $Y$ is given as
\[
Y_t=\calE_t(V)\l\{U_0+\int_{0+}^t\calE_u(V)^{-1}(dU_u-d[U,V]_u)\r\},
\]
which is much more complicated than ordinary stochastic exponentials.
As a result, a recursive calculation for a discretization of $Y$ is involved in contrast to (\ref{eq-calE}),
which means that feedback-form expressions are not appropriate to develop an approximation method for $\tvtH$.
\end{rem}

\begin{rem}\label{rem-quad}
It is almost impossible to develop a similar approximation method to (\ref{approx}) using the closed-form expression obtained by \cite{HKK},
since their expression is given as a general stochastic exponential, and includes a stochastic integral with respect to the quadratic variation of $S$,
which we cannot observe directly.
\end{rem}

\subsection{Numerical results}
We focus on the case where the process $\log(S/S_0)$($=:L^S$), is given as a variance Gamma process, and $H$ is a call option $H=(S_T-K)^+$ with $K>0$.
Note that a variance Gamma process is defined as a time-changed Brownian motion subject to a gamma subordinator.
In summary, $L^S$ is represented as
\begin{align*}
L^S_t=mG_t+\delta B_{G_t}\ \ \mbox{ for }t\in[0,T]\,,
\end{align*}
where $\delta>0$, $m\in\bbR$, $B$ is a one-dimensional standard Brownian motion, and $G_t$ is a gamma process with parameters $(1/\kappa, 1/\kappa)$ for $\kappa>0$.
Its L\'evy measure is then given as
\begin{align*}
\nu(dz)= C\l(\mathbf{1}_{\{z<0\}}e^{-G|z|}+\mathbf{1}_{\{z>0\}}e^{-M|z|}\r)\frac{dz}{|z|},
\end{align*}
where
\begin{align*}
C :=\frac{1}{\kappa} >0, \quad
G :=\frac{1}{\delta^2}\sqrt{m^2+\frac{2\delta^2}{\kappa}}+\frac{m}{\delta^2} >0, 
\quad
M :=\frac{1}{\delta^2}\sqrt{m^2+\frac{2\delta^2}{\kappa}}-\frac{m}{\delta^2} > 0. 
\end{align*}
Note that $L^S$ has no Brownian component, that is, $\sigma$ in Example \ref{ex2} is given by $0$.
As a result, the approximation (\ref{approx}) for $\tvtH_t$ simplifies to
\[
\tvtH_t\approx\txiH_t+\frac{\mu^S\calE_{t_n}}{S_{t_n}\Gamma}\sum_{k=1}^n\frac{\Delta H_{t_k}-\txiH_{t_k}\Delta S_{t_k}}{\calE_{t_k}},
\]
where
\[
\mu^S=\int_{\bbR_0}(e^z-1)\nu(dz), \ \ \ \txiH_{t_k}=\frac{K_{t_k}}{S_{t_{k-1}}\Gamma} \ \ \ \mbox{and} \ \ \
\calE_{t_{k+1}}=\calE_{t_k}\l\{1-\frac{\mu^S\Delta S_{t_{k+1}}}{S_{t_k}\Gamma}\r\}.
\]
Recall that $H_{t_k}$ and $K_{t_k}$ are computed with the FFT-based scheme developed in \cite{AIS}.

\begin{figure}[htbp]
\begin{center}
   \includegraphics[width=100mm]{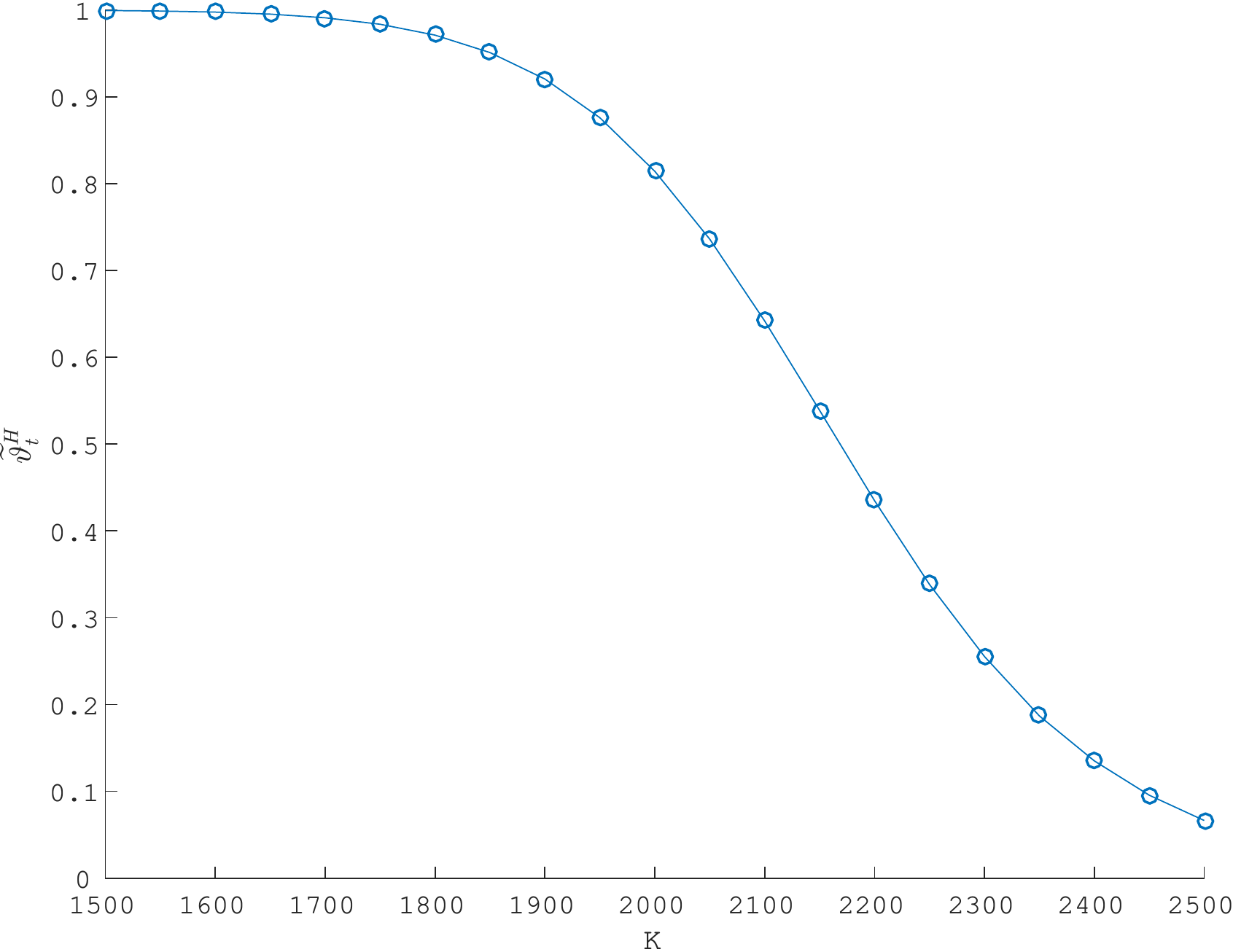}
\end{center}
\caption{Values of $\tvtH_t$ of call options at $t=121/250$ vs.\ strike price $K$ from 1500 to 2500 at steps of 50.}
\label{fig1}
\end{figure}

\begin{figure}[htbp]
\begin{center}
   \includegraphics[width=100mm]{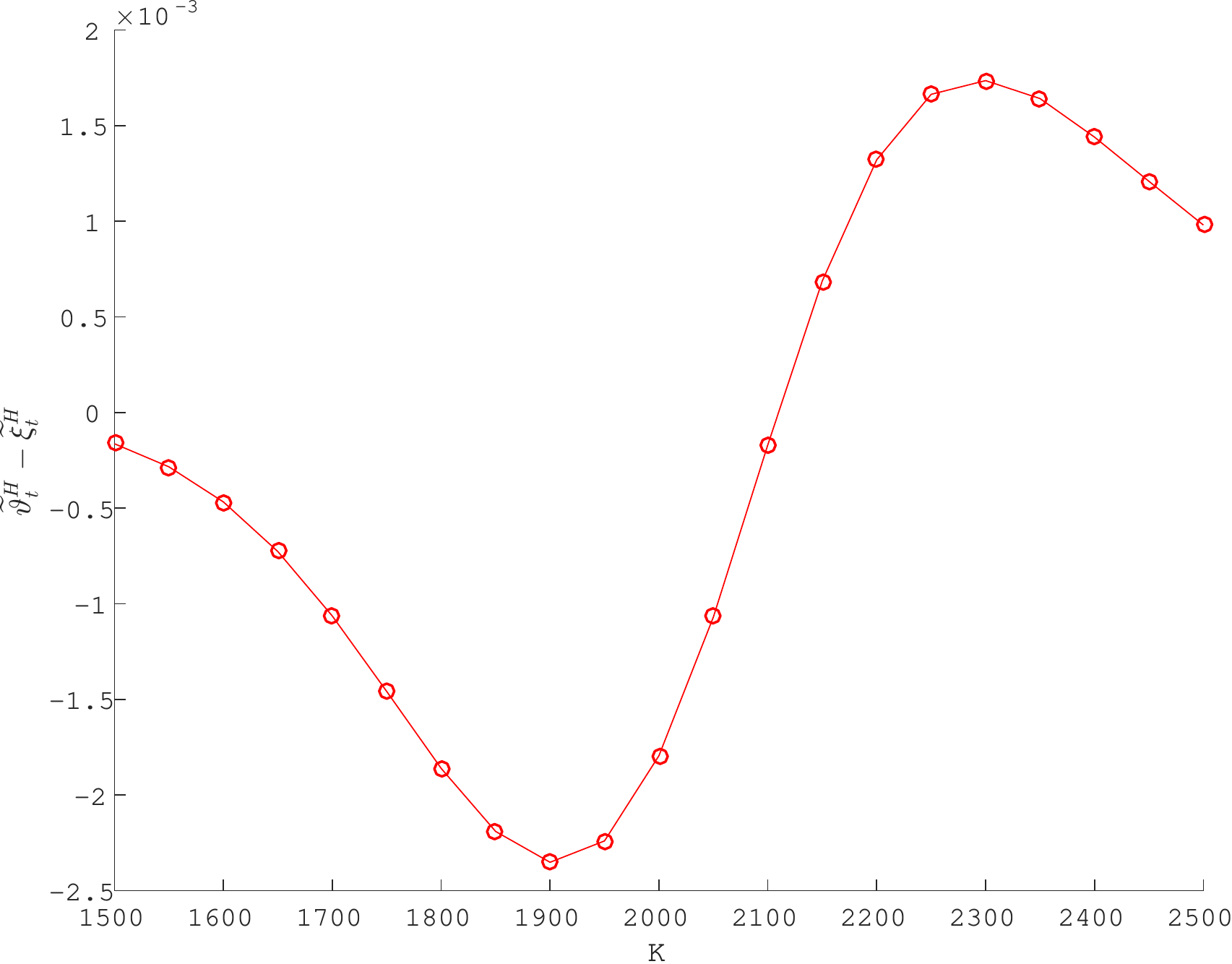}
\end{center}
\caption{Values of $\tvtH_t-\txiH_t$ vs.\ strike price $K$.}
\label{fig2}
\end{figure}

We consider European call options on the S\&P 500 Index matured on 19 May 2017, and set the initial date of our hedging to 20 May 2016.
We fix $T$ to $1$.
There are 251 business days on and after 20 May 2016 until and including 19 May 2017.
Thus, for example, 20 May 2016 and 23 May 2016 are corresponding to time $0$ and $1/250$, respectively, since 20 May 2016 is Friday.
We compute the values of MVH strategies $\tvtH$ on 10 November 2016.
Since 10 November 2016 is the 121st business day after 20 May 2016, letting $t=121/250$, we compute the values of $\tvtH_t$.
Remark that $\tvtH_t$ is constructed on 9 November 2016.
Thus, it is computed using 121 dairy closing prices of S\&P 500 index on and after 20 May 2016 until and including 9 November 2016 as discrete observational data.
As contingent claims to hedge, we consider call options with strike price 1500, 1550, $\dots$, 2500.
In addition, we set model parameters as $C=6.7910$, $G=30.1807$, and $M=33.1507$, which are calibrated by the data set of European call options on the S\&P 500 Index at 20 April 2016.
Note that the above parameter set was used in Arai and Imai \cite{AI}, and satisfies Assumptions \ref{ass1} and \ref{ass2}.
Figure \ref{fig1} shows the values of $\tvtH_t$.
The computation time to obtain the 21 values of $\tvtH_t$ on Figure \ref{fig1} is 28.85 s, which indicates that we achieve fast computation using an FFT-based method,
nevertheless computation for MVH strategies is time-consuming in general.
In addition, we compute the values of $\tvtH_t-\txiH_t$, which is the difference between the values of the MVH and LRM strategies.
Figure \ref{fig2} shows that the differences are very small, more precisely, the absolute values of $\tvtH_t-\txiH_t$ are no more than 0.0025.
Note that our numerical experiments are carried out using MATLAB (9.0.0.341360 R2016a) on an Intel Core i7 3.4~GHz CPU with 16 GB 1333 MHz DDR3 memory.

\break

\begin{center}
{\bf Acknowledgements}
\end{center}
This work was supported by JSPS KAKENHI Grant Number 15K04936.



\begin{thebibliography}{00}
\bibitem{A05}
   T. Arai, An extension of mean-variance hedging to the discontinuous case, Finance Stoch., 9 (2005), pp.129--139.
\bibitem{AI}
   T. Arai and Y. Imai, On the difference between locally risk-minimizing and delta hedging strategies for exponential L\'evy models, 
   Japan J. Indust. Appl. Math., 34 (2017), pp.845–858.
\bibitem{AIS}
   T. Arai, Y. Imai and R. Suzuki, Numerical analysis on local risk-minimization for exponential Levy models, Int. J. Theor. Appl. Finance, 19 (2016), 1650008.
\bibitem{AS}
   T. Arai and R. Suzuki, Local risk-minimization for Levy markets, Int. J. Financ. Eng., 2 (2015), 1550015.
\bibitem{B}
   F. Benth, G. Di Nunno, A. L\o kka, B. \O ksendal and F. Proske, Explicit representation of the minimal variance portfolio in markets driven by L\'evy processes,
   Math. Finance, 13 (2003), pp.55--72.
\bibitem{CK}
   A. \v{C}ern\'{y} and J. Kallsen, On the structure of general mean-variance hedging strategies, Ann. Probab., 35 (2007), pp.1479--1531.
\bibitem{CKS}
   T. Choulli, L. Krawczyk and C. Stricker, $\calE$-martingales and their applications in mathematical finance, Ann. Probab., 26 (1998), pp.853--876.
\bibitem{DTW}
   C. De Franco, P. Tankov and X. Warin, Numerical methods for the quadratic hedging problem in Markov models with jumps, J. Comput. Finance, 19 (2015), pp.29--67.
\bibitem{GOR}
   S. Goutte, N. Oudjane and F. Russo, Variance optimal hedging for continuous time additive processes and applications, Stochastics 86 (2014), pp.147-185.
\bibitem{HK}
   C. Hou and I. Karatzas, Least-squares approximation of random variables by stochastic integrals,
   in Stochastic Analysis and Related Topics in Kyoto, H. Kunita, S. Watanabe and Y. Takahashi ed., Mathematical Society of Japan, Tokyo, 2004, pp.141-166.
\bibitem{HKK}
   F. Hubalek, J. Kallsen and L. Krawczyk, Variance-optimal hedging for processes with stationary independent increments, Ann. Appl. Probab., 16 (2006), pp.853-885.
\bibitem{JS03}
   J. Jacod and A. Shiryaev, Limit Theorems for Stochastic Processes, 2nd eds., Springer, Berlin, 2003.
\bibitem{JMMS}
   M. Jeanblanc, M. Mania, M. Santacroce and M. Schweizer, Mean-variance hedging via stochastic control and BSDES for general semimartingales, Ann. Appl. Probab., 22 (2012), pp.2388-2428.
\bibitem{KV}
   J. Kallsen and R. Vierthauer, Quadratic hedging in affine stochastic volatility models, Rev. Derivatives Res., 12 (2009), pp.3-27.
\bibitem{Lim}
   A.E.B. Lim, Mean-variance hedging when there are jumps, SIAM J. Control Optim., 44 (2006), pp.1893-1922.
\bibitem{P}
   P. Protter, Stochastic Integration and Differential Equations, 2nd eds, Springer, Berlin, 2004.
\bibitem{Sch0}
   M. Schweizer, Approximation pricing and the variance-optimal martingale measure, Ann. Probab., 24 (1996), pp.206-236.
\bibitem{Sch}
   M. Schweizer, A Guided Tour through Quadratic Hedging Approaches, in Handbooks in Mathematical Finance: Option Pricing, Interest Rates and Risk Management,
   E. Jouini, J. Cvitanic and M. Musiela ed., Cambridge University Press, Cambridge, 2001, pp.538--574.
\bibitem{Sch3}
   M. Schweizer, Local Risk-Minimization for Multidimensional Assets and Payment Streams, Banach Center Publ., 83 (2008), pp.213--229.
\bibitem{S07}
   J. L. Sol\'e, F. Utzet and J. Vives, Canonical L\'evy process and Malliavin calculus, Stochastic Process. Appl., 117 (2007), pp.165--187.
\end{thebibliography}
\end{document}